\documentclass[11pt,a4paper]{article}

\usepackage[utf8]{inputenc}
\usepackage[T1]{fontenc}
\usepackage{lmodern}
\usepackage[a4paper,margin=2.3cm,headheight=14pt,footskip=28pt]{geometry}
\usepackage{amsmath,amssymb}
\usepackage{graphicx}
\usepackage{placeins}
\usepackage{natbib}
\bibpunct{(}{)}{;}{a}{}{,}
\usepackage{microtype}
\usepackage[hidelinks]{hyperref}
\usepackage{fancyhdr}

\newcommand{\publicationnote}{%
  Published in \textit{Astronomy \& Astrophysics} \textbf{713}, A34 (2026)\;--\;
  DOI: \href{https://doi.org/10.1051/0004-6361/202659810}{10.1051/0004-6361/202659810}%
}

\fancypagestyle{plain}{%
  \fancyhf{}
  \fancyfoot[C]{\footnotesize\itshape \publicationnote}
  \fancyfoot[R]{\footnotesize \thepage}

}

\title{NICER constraints on low density matching and high density continuation in neutron star equations of state}

\author{%
Federico Nola\\[0.6em]
\small Dipartimento di Matematica e Fisica, Universit\`a degli Studi della Campania ``Luigi Vanvitelli'',\\
\small viale Abramo Lincoln 5, I-81100 Caserta, Italy\\
\small Istituto Nazionale di Fisica Nucleare, Sezione di Napoli, Strada Comunale Cinthia, 80126 Napoli, Italy\\
\small Istituto Nazionale di Fisica Nucleare, Laboratori Nazionali di Frascati, C.P. 13, 00044 Frascati, Italy\\[0.4em]
\small \texttt{federico.nola@unicampania.it}
}

\date{Received March 11, 2026}

\begin{document}

\maketitle

\begin{abstract}
\noindent\textbf{Context.} Multiregion constructions of the neutron star equation of state (EoS) introduce freedom in matching low and intermediate density descriptions. This freedom can generate families of formally admissible stellar models whose physical viability must be assessed against astrophysical observations.\\[0.35em]
\textbf{Aims.} We investigated whether current neutron star constraints are sensitive not only to the high density continuation of the EoS, but also to the low density matching procedure itself. To this end, we compared two baseline constructions propagated through a common high density extension.\\[0.35em]
\textbf{Methods.} In branch $\mathcal{A}$, the sub-$n_1$ EoS was obtained by smoothly matching a low density $\beta$-equilibrated input EoS to a $\beta$-stable chiral effective field theory ($\chi$EFT) branch. In branch $\mathcal{B}$, the low density branch was continued directly up to $n_1=0.32~\mathrm{fm}^{-3}$ and served as a control construction. Above $n_1$, both branches were extended using the same parameterized continuation and were confronted with direct NICER mass--radius posteriors, a hard lower bound on the maximum mass, and an effective constraint on $\Lambda_{1.4}$. This common continuation was used to isolate the role of the low density matching sector.\\[0.35em]
\textbf{Results.} The posterior-favored predictions of branches $\mathcal{A}$ and $\mathcal{B}$ remain strongly overlapping at the observable level. The inferred stellar properties differ only slightly, and the two branches yield very similar continuation parameters, especially $n_2$ and $\Gamma_1$. Nevertheless, the NICER-informed posterior induces a nontrivial constraint on the matching parameters of branch $\mathcal{A}$, favoring $n_m \simeq 0.18~\mathrm{fm}^{-3}$ and $\Delta_m \simeq 0.021~\mathrm{fm}^{-3}$.\\[0.35em]
\textbf{Conclusions.} Current astrophysical data constrain the shared continuation above $n_1$, while indirectly restricting the low density matching sector. The matching is not arbitrary: even nearly degenerate mass--radius predictions can lead the posterior to favor matching regions. This highlights the need to constrain modular EoS frameworks observationally, where density regimes admit multiple valid connections.
\end{abstract}

\noindent\textbf{Keywords:} Dense Matter -- Equation of State -- Neutron Stars

\vspace{1em}
\section{Introduction}
\label{sec:intro}
Neutron stars (NSs) provide some of the most stringent laboratories for the physics of strongly interacting matter under extreme conditions. Their internal structure is governed by the equation of state (EoS) of cold charge-neutral matter in $\beta$-equilibrium, and therefore directly depends on the behavior of dense matter over a broad density range extending from a relatively low density crust to several times nuclear saturation density. In recent years, multimessenger observations have significantly sharpened the empirical constraints on the NS EoS. In particular, the detection of massive pulsars with masses close to or above $2\,M_\odot$, the gravitational wave signal from binary NS mergers, and X-ray pulse-profile modeling with the Neutron Star Interior
Composition Explorer (NICER) have all contributed to narrowing the allowed region in the mass--radius and tidal-deformability planes \citep{antoniadis2013massive,fonseca2021refined,abbott2019gw170817properties,kini2026j0030,dittmann2024j0740}.
Recent NICER analyses of PSR J0030+0451 and PSR J0740+6620 have provided increasingly precise information on the NS mass--radius relation \citep{Riley2019,Raaijmakers2019,kini2026j0030,dittmann2024j0740}.

At low and moderately supranuclear densities, chiral effective field theory ($\chi$EFT) provides the most systematically controlled microscopic description currently available for nucleonic matter. However, the density range relevant for NS cores extends beyond the domain in which a purely microscopic $\chi$EFT treatment can be trusted quantitatively. For this reason, practical EoS studies often combine a low density microscopic or phenomenological anchor with a parameterized continuation toward higher density. The key question is then not only how to construct a thermodynamically consistent EoS, but also which parts of the construction are actually constrained by the present astrophysical observables.

In this work, we focused specifically on the statistical status of the low density matching procedure. Rather than comparing two fully distinct EoS-building strategies, we constructed two alternative baseline branches below
\begin{equation*}
n_1 = 0.32~\mathrm{fm}^{-3},
\end{equation*}
and attached the same parameterized high density continuation to both of them. The first branch, denoted $\mathcal{A}$, was obtained by smoothly matching a low density $\beta$-equilibrated input EoS to a $\beta$-stable $\chi$EFT branch in the overlap region. The second branch, denoted $\mathcal{B}$, was continued from
the same low density input directly up to $n_1$, and therefore served as a control construction in which the explicit $\chi$EFT anchoring was removed. Above $n_1$, both branches were extended with the same continuation, consisting of a polytropic segment up to a transition density $n_2$ followed by a causal speed of sound branch.

This setup is meant to answer two related questions: whether the present data can constrain the matching procedure itself, namely the matching scale and width of the low density transition in branch $\mathcal{A}$, and how this constraint compares with the one acting on the common continuation above $n_1$. The goal is therefore not only to compare the global stellar predictions of the two branches, but also to determine whether the low density matching sector is effectively selected by the data. This controlled setup is deliberate. By keeping the high density continuation identical in the two branches, we isolate the observable imprint of the low density matching sector without attempting to determine the microscopic composition of the core.

Establishing such constraints is important because the matching region determines how uncertainties in the low density sector propagate into the full stellar sequence. Quantifying its allowed range is therefore a necessary step toward more realistic multilayer constructions, in which the crust, $\chi$EFT, and composition-dependent core EoS are matched in a statistically controlled way.

To do so, we combined a coarse Latin-hypercube \citep{McKay1979,Stein1987,MorrisMitchell1995} exploration of parameter space with a posterior-guided refinement and a likelihood built from a hard massive pulsar lower bound, together with direct NICER mass--radius posteriors for PSR J0030+0451 and PSR J0740+6620 \citep{Kini2026Data, Dittmann2024Data}. We also included the tidal sector through the quadrupolar Love number and the deformability $\Lambda_{1.4}$. For each candidate EoS, we solved the stellar structure problem and evaluated the stable mass--radius sequence, so that the NICER likelihood acts directly on the full theoretical curves.

Our main result is that current astrophysical information induces a nontrivial posterior constraint on the matching parameters of branch $\mathcal{A}$, thus showing that the low density matching procedure is statistically tested by the data. At the same time, the strongest observational selection still acts on the shared intermediate density continuation, especially through $n_2$ and $\Gamma_1$, while the matched and control branches remain strongly overlapping at the level of the main neutron-star observables.

The paper is organized as follows. In Sect.~\ref{sec:methods} we describe the construction of the two baseline branches, the common high density continuation, the stellar structure observables, and the inference pipeline. In Sect.~\ref{sec:results} we present the posterior constraints on the matching and continuation parameters, together with the resulting thermodynamic and mass--radius predictions. Section~\ref{sec:discussion} summarizes the physical implications of the analysis and outlines possible extensions.

\section{\label{sec:methods}Methods}

Our analysis was organized in four steps. We began with tabulated $\beta$-equilibrated EoSs and constructed the low density baseline. We then defined two alternative branches below $n_1$, extended both with the same high density continuation, and finally inferred the allowed parameter region from the resulting stellar structure and tidal observables.

\subsection{\label{sec:input_data}Input tables}
The numerical input consists of two tabulated EoSs generated with the MUSES Calculation Engine (MUSES CE; \citealt{PhysRevD.111.103037,MusesCalculationEngine}), a composable workflow management system that performs EoS calculations by combining models valid in complementary density regimes.
At low densities, the Crust Density Functional Theory (Crust-DFT) module provides a phenomenological description of nuclei in equilibrium with neutrons and protons in a low temperature regime. The implementation is based on the work of Lim and Holt \citep{lim2017structure}, where extended Skyrme mean-field functionals ($Sk\chi$414/450/500) are calibrated simultaneously to the bulk EoS from $\chi$EFT and to the ground-state energies of doubly magic nuclei. These calibrated functionals are then used to describe NS crust matter and, in particular, the crust--core transition.

The crust is constructed within a Wigner--Seitz framework, while the transition to uniform $\beta$-equilibrated matter is identified by comparing the energy density of the inhomogeneous configuration with that of uniform matter. Since a realistic treatment of the inner crust requires the free neutron gas to be included self-consistently, the calculation is supplemented by a compressible liquid-drop description. The resulting crust--core boundary is further checked against a thermodynamic instability criterion based on density fluctuations \citep{lim2017structure}.

Around saturation density, MUSES CE employs $\chi$EFT, which provides a systematically improvable description of nuclear matter consistent with the symmetries of low energy QCD, with nucleons and pions as the relevant degrees of freedom \citep{machleidt2011chiral}. In the present work, the microscopic input is based on a charge-dependent two-nucleon interaction constructed up to $\mathrm{N}^3\mathrm{LO}$, supplemented by three-nucleon forces at $\mathrm{N}^2\mathrm{LO}$ \citep{EntemMachleidt2003,machleidt2011chiral}. High momentum components are regulated through a smooth nonlocal cutoff,
\begin{equation*}
    f(p',p)=\exp\!\left[-\left(\frac{p}{\Lambda}\right)^{2n}-\left(\frac{p'}{\Lambda}\right)^{2n}\right],
\end{equation*}
with $\Lambda=450~\mathrm{MeV}$ and $n=3$.

The calculation is performed in many-body perturbation theory, where the three-nucleon interaction is incorporated through the normal-ordered density-dependent in-medium two-body interaction and combined with the free-space $V_{NN}$ contribution \citep{HoltKaiserWeise2009,HoltKaiserWeise2010,CoraggioHoltItacoMachleidtMarcucciSammarruca2014}. In practice, the inclusion of three-body forces is essential for obtaining a realistic pressure and density profile for NS matter; calculations based on two-body forces alone do not provide a satisfactory description of the EoS in the density range relevant for stellar structure.

In the present work, we explicitly used the baryon density $n_B$, the energy density $\varepsilon$, and the pressure $P$. The tabulated quantities are given in units of fm$^{-3}$ for number densities and MeV\,fm$^{-3}$ for $\varepsilon$ and $P$.

The low density table was truncated at
\begin{equation*}
n_B \le n_1,\qquad n_1 = 0.32~\mathrm{fm}^{-3},
\end{equation*}
while the $\chi$EFT table was used in the overlap interval
\begin{equation*}
0.032 \le n_B \le 0.32~\mathrm{fm}^{-3}.
\end{equation*}
In the actual tabulated dataset the low density branch terminates slightly below $0.32~\mathrm{fm}^{-3}$, and the final attachment to $n_1$ is therefore obtained through the same monotone interpolation used for the rest of the table.
The two inputs are not treated as temperature-dependent EoS. Although the low density file carries a nominal value $T=0.1$ MeV and the $\chi$EFT table is at $T=0$, both are regarded as effectively cold on the scale relevant for NS structure. 
To avoid spurious oscillations, the functions
\begin{equation*}
\varepsilon(n_B),\qquad P(n_B)
\end{equation*}
are interpolated with monotone piecewise cubic Hermite (PCHIP) interpolants \citep{FritschCarlson1980,FritschButland1984}.

\subsection{\label{sec:analysis_compact}EoS construction and statistical inference}

Two alternative sub-$n_1$ baseline constructions were then defined. In branch $\mathcal{A}$, the low density and $\chi$EFT branches were smoothly matched \citep{PhysRevD.111.103037} through 
\begin{equation*}
W(n_B)=\frac{1}{2}\left[1+\tanh\!\left(\frac{n_B-n_m}{\Delta_m}\right)\right],
\end{equation*}
so that in the overlap region
\begin{align*}
\varepsilon_A(n_B)&=[1-W(n_B)]\,\varepsilon_{\rm low}(n_B)+W(n_B)\,\varepsilon_{\chi{\rm EFT}}(n_B),\\
P_A(n_B)&=[1-W(n_B)]\,P_{\rm low}(n_B)+W(n_B)\,P_{\chi{\rm EFT}}(n_B).
\end{align*}
Here $n_m$ sets the center of the matching region and $\Delta_m$ its width. In branch $\mathcal{B}$, by contrast, the low density branch was continued directly up to $n_1$ and served as a control construction:
\begin{equation*}
\varepsilon_B(n_B)=\varepsilon_{\rm low}(n_B),\quad
P_B(n_B)=P_{\rm low}(n_B),\quad n_B\le n_1.
\end{equation*}

Above $n_1$, both branches were extended with the same parameterized high density continuation. Between $n_1$ and a transition density $n_2$, the pressure was modeled by a polytrope \citep{bombaci2018equation},
\begin{equation*}
P(n_B)=K\,n_B^{\Gamma_1},
\end{equation*}
with corresponding thermodynamically consistent energy density
\begin{equation*}
\varepsilon(n_B)=\frac{K}{\Gamma_1-1}n_B^{\Gamma_1}+C\,n_B,
\end{equation*}
where $K$ and $C$ are fixed by continuity at $n_1$. Above $n_2$, the EoS was continued by integrating
\begin{align*}
\frac{d\varepsilon}{dn_B}&=\frac{\varepsilon+P}{n_B},\\
\frac{dP}{dn_B}&=c_s^2(n_B)\frac{\varepsilon+P}{n_B},
\end{align*}
with a speed of sound profile approaching an asymptotic plateau:
\begin{equation*}
c_s^2(n_B)=c_{s,\mathrm{cap}}^2-
\left[c_{s,\mathrm{cap}}^2-c_s^2(n_2)\right]
\exp\!\left[-\frac{n_B-n_2}{w}\right].
\end{equation*}
The full parameter vectors are therefore
\begin{align*}
\theta_A&=(n_m,\Delta_m,n_2,\Gamma_1,w,c_{s,\mathrm{cap}}^2),\\
\theta_B&=(n_2,\Gamma_1,w,c_{s,\mathrm{cap}}^2).
\end{align*}

The prior ranges adopted for the two main continuation parameters are meant to cover qualitatively different cold, $\beta$-equilibrated, high density behaviors rather than selecting a single microscopic composition. In the scan we take $n_2\in[0.36,0.90]~\mathrm{fm}^{-3}$, so that the transition from the first polytropic segment to the speed of sound continuation can occur either just above $n_1$ or only at several times nuclear saturation density. The parameter $\Gamma_1$ controls the stiffness of the intermediate density polytropic segment: lower values correspond to a softer pressure increase after $n_1$, while larger values generate a stiffer EoS with larger pressure support before the causal continuation takes over. Thus, the explored range $\Gamma_1\in[1.5,5.5]$ deliberately includes both relatively soft and strongly stiff intermediate density continuations, subject to the stability, causality, maximum mass, radius, and tidal constraints imposed below.

All candidate equations of state are required to satisfy
\begin{equation*}
0<c_s^2=\frac{dP}{d\varepsilon}\le 1.
\end{equation*}

For each complete EoS we solved the Tolman-Oppenheimer-Volkoff (TOV) equations \citep{oppenheimer1939massive},
\begin{equation*}
    \begin{aligned}
        \frac{dP}{dr} &=
-\frac{G\left(\rho+P/c^2\right)\left(m+4\pi r^3P/c^2\right)}
{r\left(r-2Gm/c^2\right)},\\
\frac{dm}{dr} &= 4\pi r^2 \rho,
    \end{aligned}
\end{equation*}
and computed the corresponding stable stellar sequence, from which we extracted the maximum mass, characteristic radii, and the tidal sector following the standard Love-number formalism \citep{hinderer2008tidal,Hinderer2010}. The dimensionless tidal deformability is
\begin{equation*}
\Lambda=\frac{2}{3}k_2 C^{-5},
\qquad C=\frac{M}{R},
\end{equation*}
and the observables retained in the scan are
\begin{equation*}
M_{\max},\qquad R_{1.4},\qquad R_{2.0},\qquad k_2(1.4),\qquad \Lambda_{1.4}.
\end{equation*}

The statistical exploration was performed in two stages.
We first generated a coarse Latin-hypercube sampling (LHS) \citep{McKay1979,Stein1987,MorrisMitchell1995} of the parameter space of branches A and B and discarded models that violated stability or causality. Posterior weights were then assigned from the likelihood in a Bayesian inference framework \citep{Gelman2013,Trotta2008}. The observational likelihood combined three ingredients. First, we imposed a hard lower bound,
\begin{equation*}
M_{\max}\ge 2.0\,M_\odot,
\end{equation*}
motivated by the existence of NSs with accurately measured masses near or above $2\,M_\odot$ \citep{antoniadis2013massive,fonseca2021refined}. Second, we used public NICER mass--radius posterior samples directly in the $(M,R)$ plane, adopting the updated analyses of PSR J0030+0451 \citep{kini2026j0030,Kini2026Data} and PSR J0740+6620 \citep{dittmann2024j0740,Dittmann2024Data}. For each source, a kernel density estimate $p_{\rm src}(M,R)$ was built from the published posterior samples, and the source likelihood was profiled along the stable mass--radius curve predicted by each EoS:
\begin{equation*}
\ln \mathcal{L}_{\rm src}
=
\max_{j\in{\rm stable}}
\left[
\ln p_{\rm src}(M_j,R_j)-\ln p_{\rm src}^{\rm max}
\right].
\end{equation*}
The total NICER contribution is then
\begin{equation*}
\ln\mathcal{L}_{\rm NICER}
=
\ln\mathcal{L}_{\rm J0030}
+
\ln\mathcal{L}_{\rm J0740}.
\end{equation*}
Third, we retained an effective tidal term acting on $\Lambda_{1.4}$,
\begin{equation*}
\ln \mathcal{L}_{\Lambda}
=
-\frac{1}{2}\left(\frac{\Lambda_{1.4}-350}{120}\right)^2,
\end{equation*}
together with the broad requirement
\begin{equation*}
1\le \Lambda_{1.4}\le 2000
\end{equation*}
as a temporary proxy for current binary-merger constraints, broadly motivated by GW170817 \citep{abbott2019gw170817properties}. This term was used here only as a broad effective proxy for current binary-merger information and is not intended to replace a full likelihood in the binary tidal parameter $\tilde{\Lambda}$. The full log-likelihood is therefore
\begin{equation*}
\ln\mathcal{L}
=
\ln\mathcal{L}_{\rm NICER}
+
\ln\mathcal{L}_{\Lambda},
\end{equation*}
subject to the hard maximum-mass condition. 

Posterior weights were assigned as
\begin{equation*}
w_i\propto \exp\!\left(\ln\mathcal{L}_i-\max_j \ln\mathcal{L}_j\right),
\qquad
\sum_i w_i=1.
\end{equation*}
These weights were used to define refinement ranges, within which a second scan was performed. The refined samples were merged with the coarse scan and defined
the posterior used throughout the final analysis.

\section{\label{sec:results}Results}

We now discuss the main outcomes of this analysis. We first compare the thermodynamic structure of the posterior-favored EoS in order to show explicitly how the matched and control constructions differ below $n_1$ and how closely they align once the same high density continuation is attached. We then summarize the posterior constraints, and finally show the combined mass--radius region supported by the two branches.

\subsection{\label{sec:eos_results}Thermodynamic structure of the posterior-favored EoS}

The defining distinction between branches $\mathcal{A}$ and $\mathcal{B}$ lies entirely below
\begin{equation*}
n_1 = 0.32~\mathrm{fm}^{-3}.
\end{equation*}
Branch $\mathcal{A}$ is obtained by smoothly matching the low density baseline to the $\chi$EFT branch in the overlap region, whereas branch $\mathcal{B}$ continues the low density branch directly up to $n_1$, and therefore acts as a control construction.

\begin{figure}[!t]
    \centering
    \includegraphics[width=0.7\columnwidth]{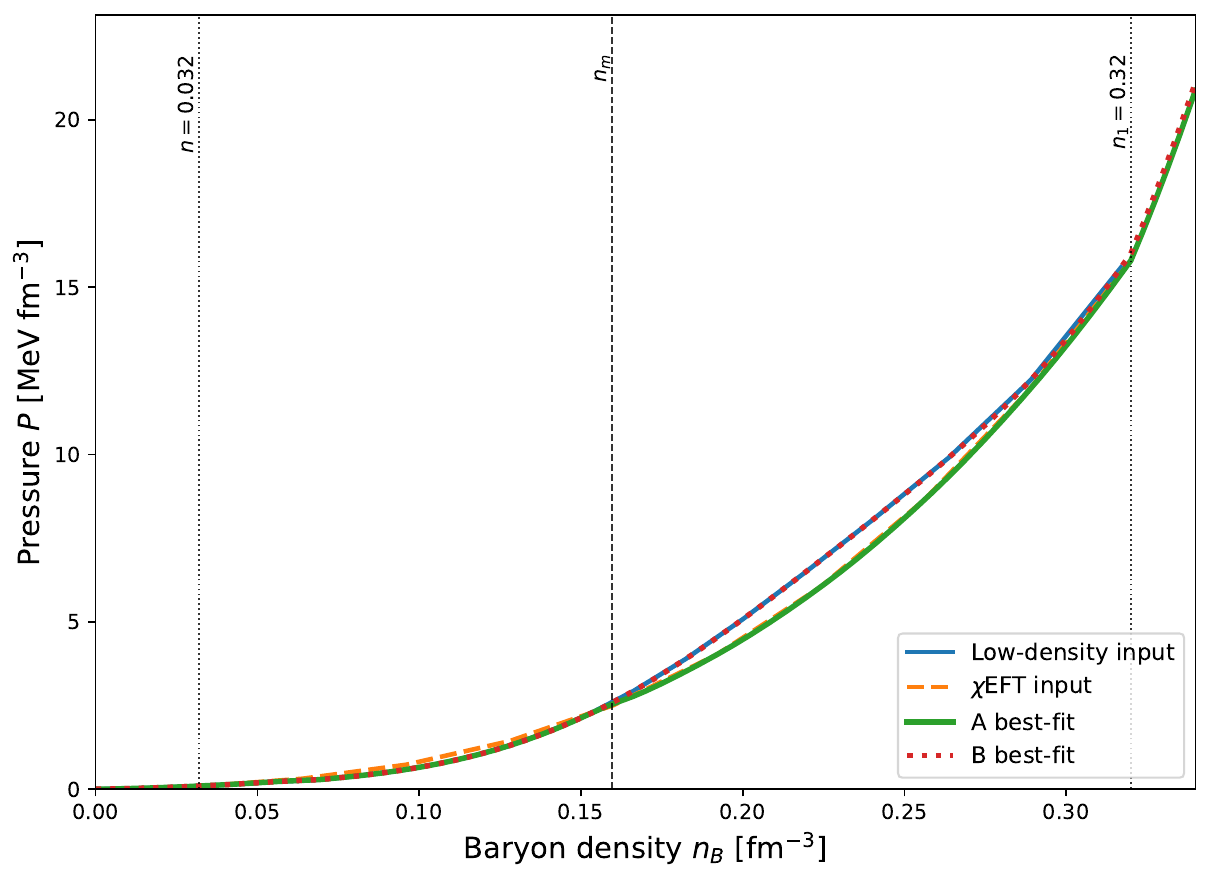}
    \caption{\label{fig:matching_region}
    Pressure as a function of baryon density in the low density and overlap region for the low density input branch, the $\chi$EFT input branch, and the best posterior representatives of branches $\mathcal{A}$ and $\mathcal{B}$. Branch $\mathcal{A}$ is obtained by smoothly matching the low density and $\chi$EFT branches through a hyperbolic tangent interpolation centered at $n_m$, while branch $\mathcal{B}$ continues the low density branch directly up to $n_1=0.32~\mathrm{fm}^{-3}$.}
\end{figure}

This difference is shown directly in Fig.~\ref{fig:matching_region}, where we plot the pressure as a function of baryon density in the low density and overlap region. As expected, the two branches differ visibly in the interval between the onset of the overlap and $n_1$: branch $\mathcal{A}$ interpolates between the low density and $\chi$EFT inputs through the matching prescription, while branch $\mathcal{B}$ retains the low density input throughout.

\begin{figure}[!t]
    \centering
    \includegraphics[width=0.7\columnwidth]{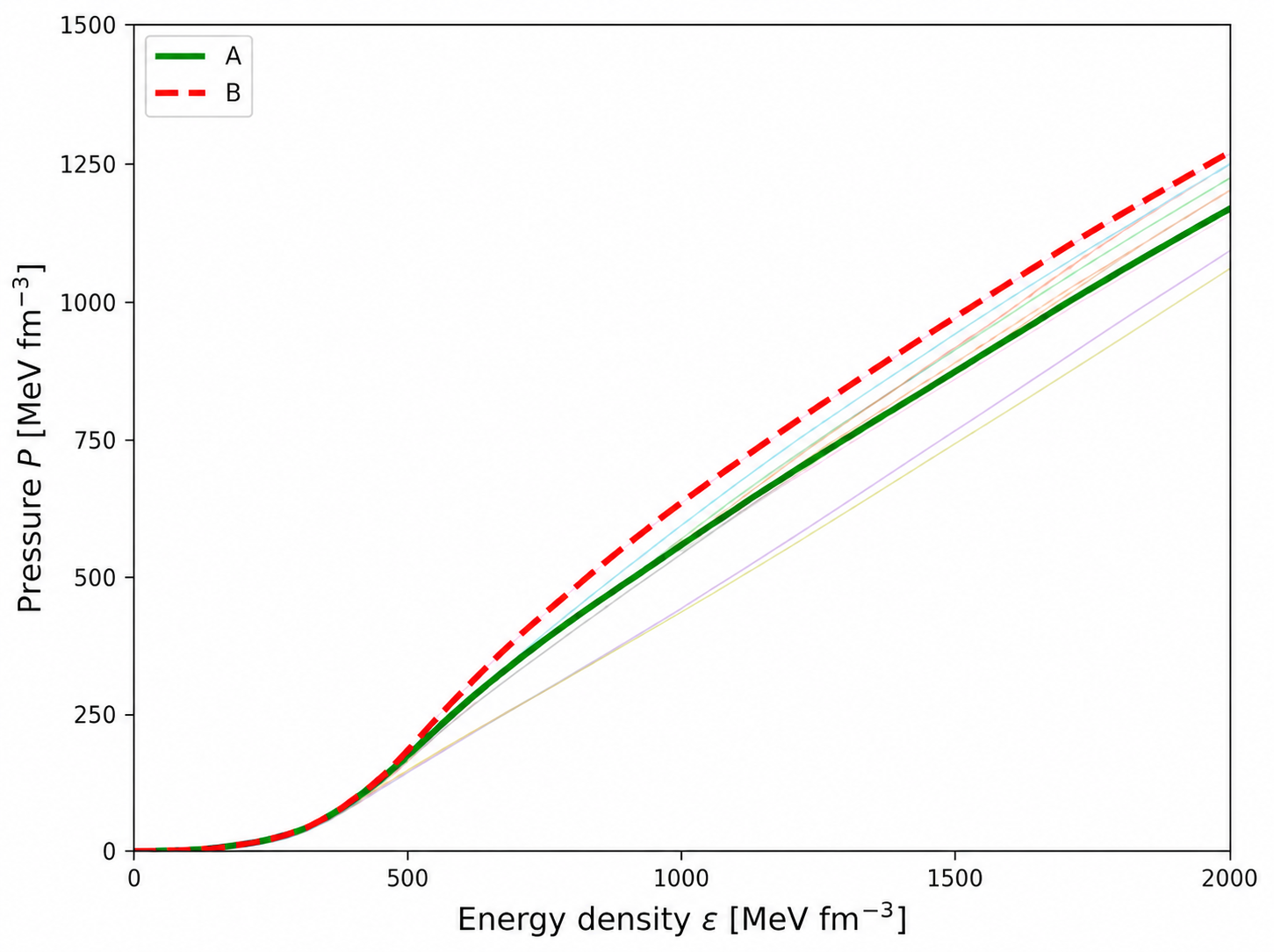}
    \caption{\label{fig:eos_peps}
    Representative posterior-favored EoS of branches $\mathcal{A}$ and $\mathcal{B}$ in the pressure--energy density plane. The thin curves denote representative high weight models, while the highlighted curves show the corresponding best posterior representatives. Once the same high density continuation is attached, the two branches occupy closely related thermodynamic regions.}
\end{figure}

The corresponding full EoS are shown in Fig.~\ref{fig:eos_peps} in the $P$--$\varepsilon$ plane. Although the two constructions differ below $n_1$, the posterior-favored full EoSs occupy closely related regions once the common high density continuation is attached. In particular, the spread that remains in the refined posterior is driven mainly by the allowed continuation parameters rather than by the distinction between the matched and control baseline constructions. This already suggests that the macroscopic stellar observables will be controlled primarily by the common continuation above $n_1$.

\subsection{\label{sec:refined_posterior_results}Posterior constraints from the refined solver-based analysis}

The initial Latin-hypercube exploration is used to identify the viable thermodynamic region and to construct posterior-guided refinement bounds. This coarse scan is then followed by a refinement step focused on the observationally preferred region, which increases the density of accepted solutions.

A central outcome of the present analysis is that the NICER-informed posterior constrains not only the common high density continuation, but also the matching sector of branch $\mathcal{A}$. In particular, the matching parameters $(n_m,\Delta_m)$ are not left arbitrary by the astrophysical likelihood. The posterior-weighted distributions yield
\begin{align*}
n_m &= 0.1798^{+0.0417}_{-0.0513}\ \mathrm{fm}^{-3},\\
\Delta_m &= 0.0207^{+0.0114}_{-0.0106}\ \mathrm{fm}^{-3},
\end{align*}
with corresponding 68\% and 95\% credible intervals
\begin{align*}
n_m &\in [0.1010,\,0.2426]\ \mathrm{fm}^{-3},\\
\Delta_m &\in [0.0054,\,0.0403]\ \mathrm{fm}^{-3}.
\end{align*}
These results show that the low density matching procedure is statistically filtered toward a restricted region of parameter space. The constraint is not extremely sharp, but it is clearly nontrivial: large portions of the prior domain are disfavored once the NICER likelihood is applied to the full stellar sequences.

\begin{figure}[!t]
    \centering
    \includegraphics[width=0.6\columnwidth]{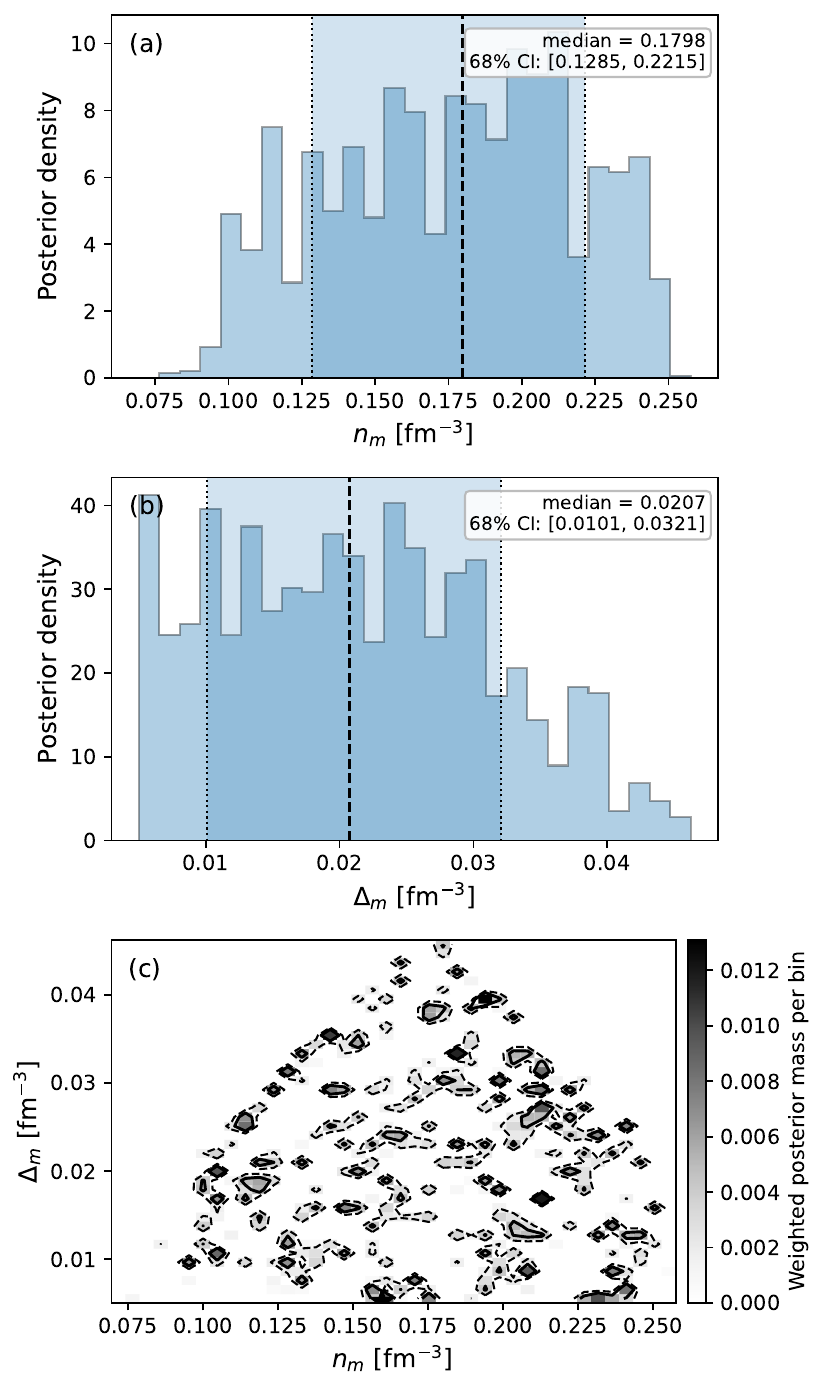}
    \caption{\label{fig:matching_params}
    Posterior-weighted constraints on the matching parameters of branch $\mathcal{A}$. The left and middle panels show the one-dimensional posterior distributions of $n_m$ and $\Delta_m$; the dashed and dotted vertical lines mark  the median and 68\% credible interval, respectively. The right panel shows the joint posterior distribution in the $(n_m,\Delta_m)$ plane, together with the 68\% and 95\% credible regions. The astrophysical likelihood excludes large portions of the prior domain, indicating a nontrivial statistical constraint on the matching sector.}
\end{figure}

This behavior is illustrated in Fig.~\ref{fig:matching_params}, where we show the posterior-weighted one-dimensional distributions of $n_m$ and $\Delta_m$, together with their joint credible region. The data therefore provide an effective indirect test of the matching procedure itself, even though the dominant observational selection still acts on the common continuation above $n_1$. This point is especially important in a modular framework such as the MUSES Calculation Engine, where different admissible matching choices can generate a broad family of formally acceptable equations of state, and consequently many mass--radius relations. The posterior constraint on $(n_m,\Delta_m)$ therefore acts as a statistical criterion for selecting which low density connections remain compatible with the data.

\begin{table}[t]
\caption{Constraints on the free parameters of branch $\mathcal{A}$.}
    \centering
    \begin{tabular}{lcc}
    \hline\hline
    Parameter  & 68\% CI & 95\% CI \\
    \hline
    $n_m~[\mathrm{fm}^{-3}]$  & [0.1285, 0.2215] & [0.1010, 0.2426] \\
    $\Delta_m~[\mathrm{fm}^{-3}]$ &  [0.0101, 0.0321] & [0.0054, 0.0403] \\
    $n_2~[\mathrm{fm}^{-3}]$ &  [0.4021, 0.6000] & [0.3701, 0.7071] \\
    $\Gamma_1$ & [3.5974, 4.6588] & [3.2797, 5.1808] \\
    $w$  & [0.2173, 0.8389] & [0.0880, 0.9761] \\
    $c_{s,\mathrm{cap}}^2$  & [0.6102, 0.9099] & [0.5191, 0.9870] \\
    \hline\hline
    \end{tabular}
    \label{tab:params_A}
\end{table}

\begin{table}[t]
\caption{Constraints on the free parameters of branch $\mathcal{B}$.}
    \centering
    \begin{tabular}{lcc}
    \hline\hline
    Parameter  & 68\% CI & 95\% CI \\
    \hline
    $n_2~[\mathrm{fm}^{-3}]$  & [0.4039, 0.5924] & [0.3696, 0.6996] \\
    $\Gamma_1$  & [3.5409, 4.7347] & [3.2252, 5.1868] \\
    $w$  & [0.1968, 0.8408] & [0.0645, 0.9802] \\
    $c_{s,\mathrm{cap}}^2$  & [0.5042, 0.8734] & [0.4305, 0.9488] \\
    \hline\hline
    \end{tabular}
    \label{tab:params_B}
\end{table}

The full posterior-weighted parameter constraints are collected in Tables~\ref{tab:params_A} and \ref{tab:params_B}. For branch $\mathcal{A}$, the common continuation parameters are constrained around
\begin{align*}
n_2 &= 0.4846^{+0.1154}_{-0.0825}\ \mathrm{fm}^{-3},\\
\Gamma_1 &= 4.0271^{+0.6317}_{-0.4297},
\end{align*}
while
\begin{align*}
w &= 0.5260^{+0.3129}_{-0.3087},\\
c_{s,\mathrm{cap}}^2 &= 0.7673^{+0.1426}_{-0.1571},
\end{align*}
with 68\% credible intervals. For branch $\mathcal{B}$ we obtain
\begin{align*}
n_2 &= 0.4752^{+0.1172}_{-0.0713}\ \mathrm{fm}^{-3},\\
\Gamma_1 &= 4.0568^{+0.6779}_{-0.5159},
\end{align*}
\begin{align*}
w &= 0.5244^{+0.3164}_{-0.3276},\\
c_{s,\mathrm{cap}}^2 &= 0.7114^{+0.1620}_{-0.2072}.
\end{align*}
The strongest overlap between the two branches is found in $n_2$ and $\Gamma_1$, confirming that the observational likelihood acts primarily on the shared intermediate density continuation. By contrast, $w$ remains comparatively weakly constrained, while $c_{s,\mathrm{cap}}^2$ shows a broader, but still nontrivial posterior selection.

In physical terms, the posterior therefore favors an intermediate density EoS that is already rather stiff between $n_1$ and $n_2$, with $n_2$ centered around $0.48~\mathrm{fm}^{-3}$, approximately three times nuclear saturation density. This behavior corresponds to an EoS that is not extremely soft immediately above $n_1$ since it must provide enough pressure support to reach $2\,M_\odot$, but also not excessively stiff at lower core densities in order to remain compatible with the relatively compact radii preferred by the NICER likelihood. The pressure cannot remain too low above $n_1$, but the subsequent speed of sound continuation prevents the construction from becoming acausal.

\begin{figure}[!t]
    \centering
    \includegraphics[width=0.9\columnwidth]{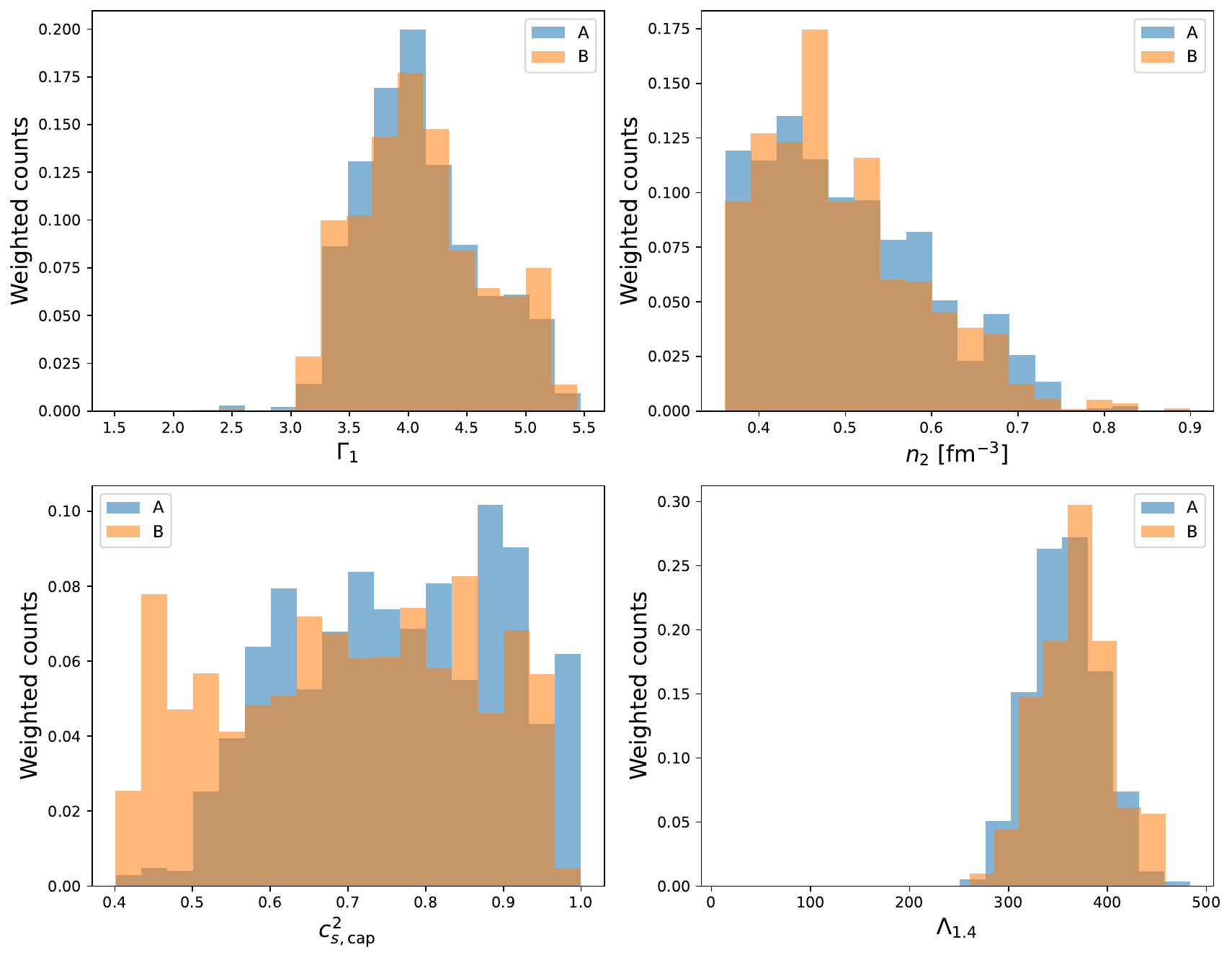}
    \caption{\label{fig:posterior_histograms}
    Posterior-weighted one-dimensional distributions for selected continuation parameters and observables in the analysis. Branches $\mathcal{A}$ and $\mathcal{B}$ remain strongly overlapping, while $\Gamma_1$ and $n_2$ continue to carry most of the constraining power.}
\end{figure}

At the observable level, the posterior-favored predictions of the two branches remain strongly overlapping once the NICER likelihood is applied directly in the $(M,R)$ plane. The differences in the posterior means of $M_{\max}$, $R_{1.4}$, $R_{2.0}$, and $\Lambda_{1.4}$ are all small. This suggests that the NICER data primarily constrain the common intermediate and high density continuation, while the difference between the matched and control constructions below $n_1$ has only a secondary effect on the global NS observables.
This pattern is also visible in the posterior-weighted one-dimensional distributions shown in Fig.~\ref{fig:posterior_histograms}, where the overlap is especially strong for $\Gamma_1$, $n_2$, and $\Lambda_{1.4}$, while the broader spread in $c_{s,\mathrm{cap}}^2$ confirms that the asymptotic part of the speed of sound branch remains less tightly constrained by the present data.

\subsection{\label{sec:mr_results}Mass--radius region}

\begin{figure}[!t]
    \centering
    \includegraphics[width=0.8\columnwidth]{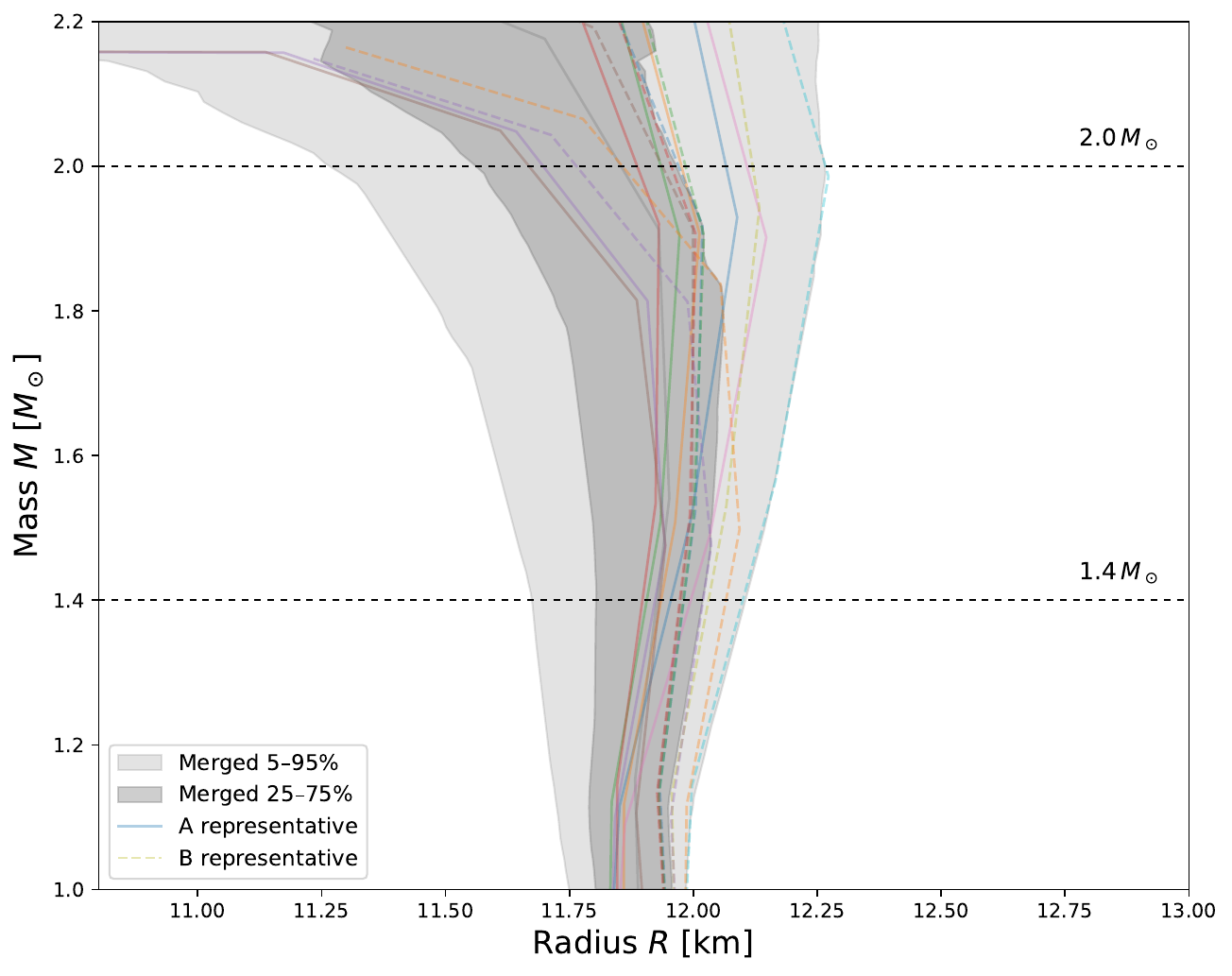}
    \caption{\label{fig:mr_bands_merged}
    Merged mass--radius credible regions obtained from the posterior-favored models of branches $\mathcal{A}$ and $\mathcal{B}$. The shaded regions denote the combined 5--95\% and 25--75\% envelopes, while the thin solid and dashed curves show representative stellar sequences from branches $\mathcal{A}$ and $\mathcal{B}$, respectively. The horizontal dashed lines mark $1.4\,M_\odot$ and $2.0\,M_\odot$.}
\end{figure}

The clearest visual summary of the final posterior is provided by the mass--radius plane. Rather than showing the two branches separately, Fig.~\ref{fig:mr_bands_merged} displays the merged 5--95\% and 25--75\% envelopes obtained from the posterior-favored models of branches $\mathcal{A}$ and $\mathcal{B}$, together with representative stellar sequences from the two branches. This presentation highlights the total region supported by the analysis while preserving the internal spread of the allowed solutions.

Two features are immediately apparent. First, the allowed mass--radius region is compact in the canonical-mass range and remains centered around radii close to $12$ km. Second, the representative sequences from the two branches populate essentially the same region over the full astrophysically relevant mass interval shown here. In particular, the overlap persists up to and beyond the $2 M_\odot$ scale, confirming that the weak sensitivity to the sub-$n_1$ baseline construction persists when the NICER terms are applied directly to the full theoretical mass--radius curves.

\FloatBarrier

\section{\label{sec:discussion}Discussion and conclusions}

The present NICER-based analysis clarifies the main result of this
work. The most important point is that the astrophysical likelihood constrains not only the continuation above $n_1$, but also, indirectly, the matching sector of branch $\mathcal{A}$. In
particular, the parameters $(n_m,\Delta_m)$ are not free to vary arbitrarily once the full stellar sequences are compared with the NICER mass--radius posteriors.

The hierarchy of the constraints is nevertheless clear. The strongest selection still acts on the common continuation above $n_1=0.32~\mathrm{fm}^{-3}$, especially through $n_2$ and $\Gamma_1$, whose posterior values remain very similar in branches $\mathcal{A}$ and $\mathcal{B}$. By contrast, the matching parameters $(n_m,\Delta_m)$ are constrained more indirectly, through their effect on the full stellar sequence and hence on the NICER likelihood. Their posterior region is therefore broader, but not structureless.

The sensitivity to the fixed value of $n_1$ should be interpreted accordingly. In this work $n_1=0.32~\mathrm{fm}^{-3}$ is kept fixed, close to the upper end of the low density and $\chi$EFT input domain, so the analysis does not perform an independent marginalization over the attachment density itself. The main conclusions are therefore conditional on this choice. Within the present setup, however, the sensitivity to $n_1$ is expected to be subdominant with respect to the posterior selection of $n_2$ and $\Gamma_1$ because $n_1$ is common to both branches and mainly fixes where the shared continuation is attached. Nevertheless, because the two branches have the same value of $n_1$ and the same continuation above it, the branch comparison is mainly sensitive to the matching prescription below $n_1$ and to the common parameters $n_2$ and $\Gamma_1$. Varying $n_1$ would mostly redistribute the density interval assigned to the low density matching sector and to the first polytropic segment, and a dedicated scan over $n_1$ would be a useful extension of the present study.

The choice $n_1=0.32~\mathrm{fm}^{-3}$ is mainly technical. It allows us to use the full density range covered by the $\chi$EFT module available in the MUSES Calculation Engine, and it should not be interpreted as an assumption that the microscopic composition remains purely nucleonic up to or beyond this density.
In some microscopic EoS models, hyperons, $\Delta$ resonances, or other exotic degrees of freedom may appear at approximately twice the nuclear saturation density, close to the adopted value of $n_1$. Their appearance, however, is not automatic; each species must be energetically favored and must also be
compatible with charge neutrality and $\beta$-equilibrium. As a result, only a subset of the possible additional degrees of freedom may actually populate the star and their particle fractions may remain relatively small over the density range relevant for the present configurations. A related composition-dependent calculation with the MUSES Calculation Engine
illustrates this point by comparing standard nucleonic, $\Delta$-enriched, and hyperonic scenarios, showing that the resulting particle fractions depend on the imposed $\beta$-equilibrium and charge neutrality conditions
\citep{AriasAragonNola2026}. In the present analysis
these composition changes are not modeled explicitly, but their possible macroscopic effect is effectively absorbed into the generic high density continuation above $n_1$. A natural extension would therefore be to apply the same statistical framework to models in which the generic high density continuation is replaced by composition-dependent core EoSs computed within MUSES CE, including the possible appearance of different hyperonic species and/or baryonic resonances.

This picture is consistent with the strong overlap of the macroscopic predictions of the two branches. With a likelihood built from NICER mass--radius posteriors, the posterior-favored values of $M_{\max}$, $R_{1.4}$, $R_{2.0}$, and $\Lambda_{1.4}$ remain close in the two branches. The data therefore do test the low density matching sector, but the dominant observational selection still falls on the shared intermediate and high density continuation.

The present framework is not intended to identify the microscopic composition of the stellar core. Its purpose is narrower: to determine which parts of a modular multiregion construction are already constrained by current data, and in particular whether the low density matching sector is statistically selected or left effectively free. This is precisely why a statistical calibration of the matching freedom is useful; it identifies which parts of the construction are already selected by current observations before one moves to more composition-specific core models.

This point is especially important in the context of MUSES CE \citep{PhysRevD.111.103037,MusesCalculationEngine}. A main strength of that framework is its modularity: descriptions that are valid in different density regimes can be combined into a single NS EoS. That same flexibility, however, also means that the matching between regions is not unique a priori. Different admissible choices in the transition between crust, $\chi$EFT, and higher density sectors can generate a large family of formally acceptable EoSs and, consequently, a broad set of mass--radius curves, not all of which are supported by the data.

For this reason, constraining the matching sector is physically important. It is the step that allows one to distinguish between the broad modeling freedom offered by a multiregion framework and the subset of constructions that remain compatible with current observations. In this sense, the present analysis reduces a technical ambiguity, and also provides a statistical criterion for selecting which low density matching prescriptions are actually viable once the full stellar sequence is confronted with the NICER posteriors. This is also the natural starting point for future MUSES CE based studies with explicit composition-dependent cores, where the number of admissible multiregion constructions will be even larger.

Two natural extensions follow from this analysis. On the observational side, the most immediate next step is to replace the present effective $\Lambda_{1.4}$ term with a full binary-merger likelihood in the tidal parameter $\tilde{\Lambda}$. On the modeling side, we are already developing a related study in which the generic high density continuation is replaced by explicit MUSES CE core EoS with different admissible compositions (standard $npe\mu$, octet-enriched, decuplet-enriched, and quark-matter sectors).

In summary, the present analysis shows that the low density matching sector is not arbitrary. In this sense, the present analysis should be viewed as a first statistical calibration of the matching freedom in a modular EoS framework. This calibration is a necessary step before one can compare more realistic composition-dependent core models on the same footing. In a framework where multiple density regions can be combined in different admissible ways, such a calibration is necessary in order to distinguish genuine observational information from residual modeling freedom. The current data constrain the shared intermediate density continuation most strongly, but they also select a restricted matching region through its imprint on the full stellar sequence. This is the central result of the paper: in a modular multiregion framework, the current data not only constrain the shared continuation, but also begin to calibrate the matching freedom itself.

\section*{Acknowledgements}
I acknowledge the use of the MUSES Calculation Engine for the NS structure calculations performed in this work. I also thank the MUSES CE developers and community for their helpful feedback and support through the official forum. 
I would like to thank my supervisor, Nunzio Itaco, for their invaluable support and guidance throughout the writing of this paper, as well as for their helpful feedback and suggestions that significantly improved the manuscript.

\bibliographystyle{plainnat} 
\bibliography{Using}

\end{document}